\documentclass{article}

\usepackage[english]{babel}

\usepackage[letterpaper,top=2cm,bottom=2cm,left=3cm,right=3cm,marginparwidth=1.75cm]{geometry}

\usepackage{amsmath,physics}
\usepackage{graphicx}
\usepackage{tikz}
\usepgflibrary {shadings}
\title{figure}
\usepackage{pgfplotstable}
\pgfplotsset{compat=1.18}
\usetikzlibrary{positioning}
\usepackage[outline]{contour}
\usepackage[colorlinks=true, allcolors=blue]{hyperref}
\usepackage{rotating}
\usepackage{pdflscape}
\usepackage{booktabs} 
\usepackage{comment}
\usepackage{footnote}

\usepackage{threeparttable}
\usepackage{caption}
\usepackage{booktabs}
\usepackage{geometry}

\usepackage{xcolor}
\usepackage{tikz}
\usetikzlibrary{positioning, shapes.geometric, arrows}
\usepackage{array}
\usepackage{rotating}  
\usepackage{pdflscape}  
\usepackage{graphicx}   
\usepackage{hyperref} 
\usepackage{cite}

\title{Quantum Teleportation 
of a Single Qutrit using Two-Qutrit Entangled States
}
\author{You}  

\author{
    Surajit Sen \textsuperscript{1},
    Tushar Kanti Dey\textsuperscript{1},
    Anushree Bhattacharjee\textsuperscript{1,2},
    Sovik Roy\textsuperscript{1,2}
    \\[1em] 
    \textsuperscript{1} Centre of Advanced Studies and Innovation Lab,\\ 18/27 Kali Mohan Road, Tarapur, Silchar 788003, INDIA \\\\
    \textsuperscript{2} Department of Mathematics, Techno Main Salt Lake (Engg. Colg.),\\ EM 4/1, Sector V,  
    Kolkata 700091, INDIA \\
}
\begin{document}
\maketitle

\renewcommand{\thefootnote}{\arabic{footnote}} 
\footnotetext[1]{Corresponding author: ssen55@yahoo.com}
\footnotetext[2]{tkdey54@gmail.com}
\footnotetext[3]{s.roy2.tmsl@ticollege.com}
\footnotetext[4]{a.bhattacharjee.tmsl@ticollege.com}

\begin{abstract}
\noindent 
We demonstrate quantum teleportation of a qutrit system using a complete set of two-qutrit entangled states obtained from the representation theory of the SU(3) group. All measurement gates essential for end-to-end teleportation are systematically evaluated, and these are found to be non-unitary. Our approach extends Bennett’s teleportation protocol to the qutrit system with minimal modifications, preserving operational simplicity and underscoring the necessity of non-unitary measurement operators in high-dimensional systems.
\end{abstract}


\section{Introduction:}
\par 
Quantum teleportation is widely recognized as one of the most significant protocols for transferring information over distances. Originally proposed by Bennett et al. in 1993 \cite{Bennett1993,Nielsen2010}, it relies on the entanglement as the operational resource which was experimentally validated through photon polarization experiment \cite{Bouwmeester1997,Zeilinger2000}, implemented for two remote atomic-ensemble of large number of rubidium atoms \cite{Bao2012,Boschi1998} and even successfully tested up to metropolitan-scale fiber networks extending to several kilometers \cite{Valivarthi2016,Shen2023}. Taking advantage of the properties inherent in quantum mechanics, entanglement enables highly secure communication channels, leading to the development of quantum cryptography \cite{Bennett1984,Nielsen2010,Gisin2002}. Today, quantum teleportation is considered one of the key mechanism for advancing the Quantum Internet, which not only aims to establish high-fidelity large-scale communication network \cite{Bao2012,Takeda2013,DeMartini2019}, but is also believed to lay the foundation of much-anticipated Quantum Internet of Things (Q-IoT) \cite{Pirandola2015,Hossain2024,Sikiru2024}. \\\\
\noindent
Before delving into quantum teleportation of a qutrit system, let us first consider the tenets of Bennett,s  teleportation protocol using two-qutrit entangled states instead of two-qubit EPR states. To enunciate how it works, let us consider an observer Alice prepares a qutrit state $\ket{\phi}_{A_1}$ in her lab `\text{$A_1$}' and ask the post office (P.O.) located at `\text{$A_2$}' to teleport it to a remote observer Bob located at `B'. The whole process requires an two-qutrit entangled state $\ket{\psi_{A_2B}}$ shared between the post office interface at `$\text{A}_2$', where teleportation was initiated and the terminal point `B', where Bob will receive it. Thereafter, to decode the information transferred, Bob needs an appropriate message in form of an entangled state $\ket{\psi_{A_1A_2}}$ from Alice-P.O. interface via a classically transmitted signal. This enables him to retrieve the information of the input state at site `B' using appropriate measurement gate (operator) $\Lambda_i$ constructed in SU(3) basis. Together, these steps constitute an augmented version of Bennett's  protocol for qutrits, which we proceed to implement for qutrit teleportation \cite{Bao2012,Takeda2013,DeMartini2019}.\\\\
\noindent 
In Hilbert space $\mathcal{H}^3$, the qutrit is characterized by the three-level system with distinct transition selection rules \cite{Hioe1982,Yoo1985,Nath2008,Sen2012}. Over the years in the quantum optics frontier, three-level systems have been extensively studied, extending the scope of conventional two-level (qubit) systems. These systems underpin a diverse range of phenomena, such as superradiance \cite{Dicke1954}, the quantum Zeno effect \cite{Misra1977}, coherent population trapping (CPT) \cite{Alzetta1976,Gray1978}, electromagnetically induced transparency (EIT) \cite{Harris1990,Boller1991}, and lasing without inversion (LWI) \cite{Harris1989}, among others \cite{Sen2015,Sen2017,Sen2023}. 
Despite these advancements, it is well-understood that the entanglement scenario for the qutrit system is not just a scaled-up version of qubit entanglement, and therefore demands new theoretical tools to address associated quantum information phenomena. In the case of the qutrit teleportation in particular, most studies primarily deal with the Maximally Entangled Singlet (MES) state as a key resource in absence of full spectrum of entangled qutrit states   \cite{Lanyon2009,PanLu2006,Luo2019,Erhard2020,Huang2020,Leslie2019,Hu2020,Roy2025}. Recently, we have developed a set of two-qutrit entangled states, the qutrit counterpart of the EPR states, utilizing the representation theory of the SU(3) group — the mathematical framework that describes how to combine quantum states \cite{Sen2024}. In this paper, we study quantum teleportation of a qutrit using this set of states within the framework of an augmented version of Bennett's protocol, as described above, and find that all possible measurement operators are non-unitary. 
\\\\ 
\noindent 
The remainder of this paper is organized as follows: In Section~II, we review the two-qutrit entangled states, the key resource necessary for qutrit teleportation. Section~III gives an outline the formalism of qutrit teleportation, drawing a parallel analogy with the standard Bennett protocol. In Section~IV, we show that even in absence of fault-tolerant unitary measurement gates for qutrits, our approach offers a consistent framework for its teleportation without requiring the unitarity of the measurement operators a priori. In the final section, we will summarize the key findings of our work.

\section{Two-qutrit entangled states revisited}
\par 
We begin by taking the input qutrit state hold by Alice at location `\text{$A_1$}',  
\begin{align} 
    \ket{\phi}_{A_1} = c_0 \ket{0}_{A_1}  + c_1 \ket{1}_{A_1} + c_2 \ket{2}_{A_1},
    \label{eq1}
\end{align}
where $c_0, c_1, c_2$ be the normalized amplitudes with the standard basis states given by, 
\begin{align}
     \ket{0}_{A_1}=\begin{pmatrix} 1 \\ 0 \\ 0 \end{pmatrix}_{A_1}\;, \quad 
     \ket{1}_{A_1}=\begin{pmatrix} 0 \\ 1 \\ 0 \end{pmatrix}_{A_1},\; \quad 
     \ket{2}_{A_1}=\begin{pmatrix} 0 \\ 0 \\ 1 \end{pmatrix}_{A_1}\;
     \label{eq2}.
\end{align}
In order to teleport the state from a post office interface located at `\text{$A_2$}' to a distant observer Bob at `B', Alice requires two-qutrit entangled states shared between `\text{$A_2$}' and `\text{$B$}'. 
Recently we have constructed a set of two-qutrit entangled states using SU(3) representation theory
\cite{Sen2024} and broadly speaking, these nine states 
are classified into following categories:
\\

i) Singlet state (Maximally Entangled State (MES)): \\ 
\begin{subequations}
\begin{align} 
\ket{\Psi_{0}}_{{A}_2B}&= \frac{1}{{\sqrt 3 }}\big[\ket{0_{A_2}}\ket{0_{B}}  + \ket{1_{A_2}}\ket{1_{B}} + \ket{2_{A_2}}\ket{2_{B}}\big], 
\label{eq3a}
\end{align}
%
ii) Bell-like states: 
\begin{align} 
\ket{\Psi_{1}}_{A_2B}&= \frac{1}{\sqrt 2}\big[ \ket{1_{A_2}}\ket{0_{B}} + \ket{0_{A_2}}\ket{1_{B}} \big],
\label{eq3b}
\end{align}
\begin{align} 
\ket{\Psi_{2}}_{A_2B}&= \frac{1}{{\sqrt 2 }}\big[\ket{1_{A_2}}\ket{0_{B}} - {\ket{0_{A_2}}\ket{1_{B}}} \big],
\label{eq3c}
\end{align}
\begin{align} 
\ket{\Psi_{3}}_{A_2B} &= \frac{1}{\sqrt 2}\big[-\ket{1_{A_2}}\ket{1_{B}} + \ket{2_{A_2}}\ket{2_{B}}\big],
\label{eq3d}
\end{align}
\begin{align} 
\ket{\Psi_{4}}_{A_2B}  &= \frac{1}{\sqrt 2}\big[ \ket{2_{A_2}}\ket{0_{B}} + \ket{0_{A_2}}\ket{2_{B}}\big],
\label{eq3e}
\end{align}
\begin{align} 
\ket{\Psi_{5}}_{A_2B} &= \frac{1}{\sqrt 2}\big[\ket{2_{A_2}}\ket{0_{B}} - \ket{0_{A_2}}\ket{2_{B}}\big],
\label{eq3e}
\end{align}
\begin{align} 
\ket{\Psi_{6}}_{A_2B}  &= \frac{1}{\sqrt 2}\big[ \ket{2_{A_2}}\ket{1_{B}}  + \ket{1_{A_2}}\ket{2_{B}}\big],
\label{eq3f}
\end{align}
\begin{align} 
\ket{\Psi_{7}}_{A_2B}  &= \frac{1}{\sqrt 2}\bigg[\ket{2_{A_2}}\ket{1_{B}} - \ket{1_{A_2}}\ket{2_{B}}\big],
\label{eq3g}
\end{align}
iii) Pure octet state: 
\begin{align} 
\ket{\Psi_{8}}_{A_2B} &= \frac{1}{\sqrt 6 }\big[ - 2\ket {0_{A_2}}\ket{0_B} + \ket{1_{A_2}}\ket{1_B} + \ket{2_{A_2}}\ket{2_B}\big].
\label{eq3h}
\end{align}
\label{eq3}
\end{subequations}
It is customary to express the computational basis states in Eq.(3) in terms of the entangled states, 
\begin{subequations}
\begin{align} 
\ket{0_{A_2}}\ket{0_B} &= \frac{1}{{\sqrt 3 }}\big[{{\ket{ {{\Psi_{0}}} } }_{A_2B}} - \sqrt 2  {{\ket{ {{\Psi _{8}}} } }_{A_2B}}\big],\\
\ket{1_{A_2}}\ket{1_B} &= \frac{1}{{\sqrt 3 }}\big[\ket{ {{\Psi_{0}}} } _{A_2B} - \sqrt{\frac{3}{2}}\ket{ {{\Psi _{3}}} } _{A_2B} + \frac{1}{{\sqrt 2 }}\ket{ {{\Psi_{8}}} }_{A_2B}\big],\\
\ket{1_{A_2}}\ket{0_B} &= \frac{1}{{\sqrt 2 }}\big[{{\ket{ {\Psi _{1}} } }_{A_2B}} + {\ket{ {\Psi _{2}} } }_{A_2B} \big],\\ 
\ket{0_{A_2}}\ket{1_B} &= \frac{1}{{\sqrt 2 }}\big[{{\ket{{\Psi_{1}} } }_{A_2B}} - {\ket{ {\Psi _{2}} } }_{A_2B} \big],\\
\ket{2_{A_2}}\ket{0_B} &= \frac{1}{{\sqrt 2 }}\big[ {{\ket{ {\Psi _{4} } } }_{A_2B}} + {\ket{ {\Psi_{5}} } }_{A_2B} \big],\\
\ket{0_{A_2}}\ket{2_B} &= \frac{1}{{\sqrt 2 }}\big[{{\ket{ {\Psi _{4} } } }_{A_2B}} - {\ket{ {\Psi _{5} } } }_{A_2B} \big]. \\
\ket{2_{A_2}}\ket{1_B} &= \frac{1}{{\sqrt 2 }}\big[{{\ket{ {\Psi _{6}} } }_{A_2B}} + {\ket{ {\Psi _{7}} } }_{A_2B} \big],\\
\ket{1_{A_2}}\ket{2_B} &= \frac{1}{{\sqrt 2 }}\big[{{\ket{ {\Psi _{6}} } }_{A_2B}} - {\ket{ {\Psi_{7} } } }_{A_2B} \big],\\
\ket{2_{A_2}}\ket{2_B} &= \frac{1}{{\sqrt 3 }}\big[\ket{ {{\Psi_{0}}} }_{A_2B} +  \sqrt{\frac{3}{2}}\ket{ {{\Psi_{3}}} } _{A_2B} + \frac{1}{{\sqrt 2 }}\ket{ {{\Psi_{8}}} }_{A_2B}\big].
\end{align}
\label{eq4}
\end{subequations} 
To implement an augmented Bennett's protocol for the qutrit system we now proceed to derive requisite non-unitary measurement gates using above relations.
 
\section{Formalism}
\par 
To facilitate teleportation of a single qutrit state $\ket{\phi}_{A_1}$ using a generic entangled state $\ket{\Psi_{i}}_{A_2B}$, we first construct a composite state from Eqs.\eqref{eq1} and \eqref{eq3},
\begin{align}
    \ket{\xi^{}_{i}}_{A_1A_2B} := \ket{\phi}_{A_1} \otimes \ket{\Psi_{i}}_{A_2B}, 
\label{eq5}
\end{align}
where $i = 0,1,2,\dots, 8$ are nine distinct channels which correspond to nine entangled states given by Eq.\eqref{eq3}. Fig. 1 illustrates the schematic diagram of qutrit teleportation using nine distinct channels. Using the associative property of tensor product, we can express Eq.(5) as the tensor product of entangled state $\ket{\Psi_{i}^k}_{A_1A_2}$, shared between Alice's lab  and her P.O. interface, a state $\ket{s_i^k}_B$ which we refer as the pre-measurement state, 

\begin{figure}[htbp]
    \centering
    \includegraphics[width=1.0\textwidth]{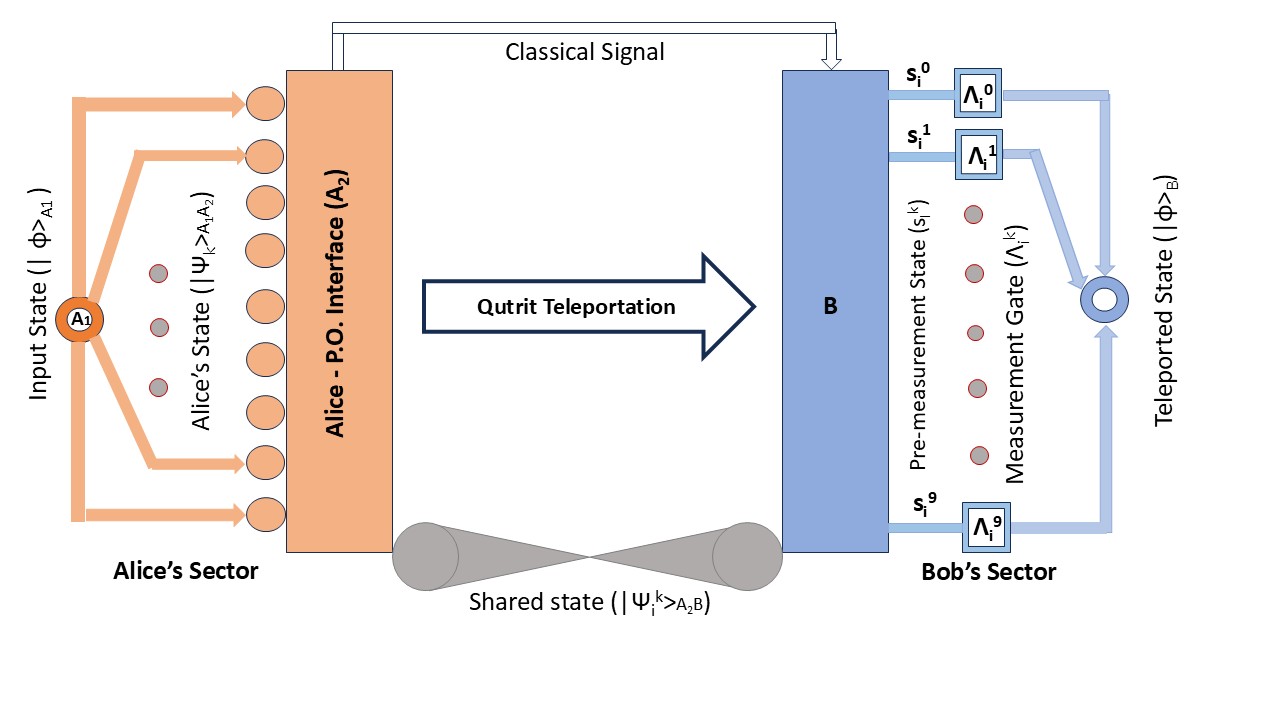} 
\caption{Schematic diagram of qutrit teleportation from $\ket{\phi}_{A_1}$ to $\ket{\phi}_{B}$ using nine two-qutrit entangled states $\ket{\Psi_{i}}_{A_2B}$ shared between Alice and Bob following an augmented version of Bennett's protocol. Here, Alice's sector contains nine states $\ket{\Psi_{k}}_{A_1A_2}$ for each channel, while Bob's sector features pre-measurement states $\ket{s_{i}^k}_B$ and non-unitary measurement operators $\Lambda^{k}_{i}$, all consistent with SU(3) symmetry.}
    \label{fig:qutrit_teleport}
\end{figure}

\begin{equation}
\begin{split}
\ket{\xi_{i}}_{A_1A_2B} &= \sum_{k=0}^{8} \ket{\Psi_{k}}_{A_1A_2} \otimes \ket{s^{k}_i}_B, \\
&= \sum_{k=0}^{8} \ket{\Psi_{k}}_{A_1A_2} \otimes \bigg\{ \big[\Lambda^{k}_{i}\big] \ket{\phi}_{B} \bigg\}.
\end{split}
\label{eq6}
\end{equation}
where $k$ represents 9-fold multiplicity of each channel. In last line of Eq.\eqref{eq6}, the state $\ket{s_{i}^k}_B$ is obtained by operating the measurement gate $\big[\Lambda^{i}_{k}\big]$ onto the qutrit state $\ket{\phi}_B$ which is relocated at `B'. That is, for $i$-th channel we have, 
\begin{align}
\ket{s^{k}_i}_B=\big[\Lambda_i^k\big]\ket{\phi}_B, 
\label{eq7}
\end{align}
which gives the desired condition of teleportation. In the following Section we use Eq.\eqref{eq7} to find all measurement gates for different communication channels.

\section{Qutrit measurement gates}
\par 
To begin, let us first illustrate the teleportation of the qutrit state $\ket{\phi}_{A_1}$ for channel-I using the entangled state $\ket{\Psi_{0}}_{A_2B}$. To achieve that, substitution of Eqs.\eqref{eq1} and \eqref{eq3a} into Eq.\eqref{eq5} yields,   
\begin{align}
    \ket{\xi_{0}}_{A_1A_2B} &= \ket{\phi}_{A_1} \otimes \ket{\Psi_{0}}_{A_2B}  \nonumber \\ 
     &=   
    \big[c_0 \ket{0}_{A_1} + c_1 \ket{1}_{A_1} + c_2 \ket{2}_{A_1}\big]\otimes  \frac{1}{{\sqrt 3 }}\big[\ket{0_{A_2}}\ket{0_{B}}  + \ket{1_{A_2}}\ket{1_{B}} + \ket{2_{A_2}}\ket{2_{B}}\big], 
\label{eq8}
\end{align}
which using Eq.(4) becomes, 
\begin{align}
\ket{\xi_{0}}_{A_1A_2B} &=\ket{\Psi_{0}}_{A_1A_2}\otimes\ket{s_{0}^0}_B \nonumber \\ 
&+\ket{\Psi_{1}}_{A_1A_2}\otimes\ket{s_{0}^1}_B
+\ket{\Psi_{2}}_{A_1A_2}\otimes\ket{s_{0}^2}_B \nonumber \\
& +\ket{\Psi_{3}}_{A_1A_2}\otimes\ket{s_{0}^3}_B \nonumber\\
&+\ket{\Psi_{4}}_{A_1A_2}\otimes\ket{s_{0}^4}_B
+\ket{\Psi_{5}^0}_{A_1A_2}\otimes\ket{s_{0}^5}_B \nonumber  \\
&+\ket{\Psi_{6}}_{A_1A_2}\otimes\ket{s_{0}^6}_B + 
\ket{\Psi_{7}}_{TA}\otimes\ket{s_{0}^7}_B \nonumber \\
&+\ket{\Psi_{8}}_{A_1A_2}\otimes\ket{s_{0}^8}_B,
\label{eq9}
\end{align}
where the pre-measurement states at `$B$' are given by,  
\begin{subequations}
\begin{align}
\ket{s_{0}^0}_B&:=\frac{1}{3}\big[c_0\ket{0_B}+c_1\ket{1_B}+c_2\ket{2_B}\big],\\
\ket{s_{0}^1}_B&=\frac{1}{\sqrt{6}}\big[c_1 \ket{0_B} + c_0\ket{1_B}\big],\\
\ket{s_{0}^2}_B&=\frac{1}{\sqrt{6}}\big[c_1 \ket{0_B}-c_0\ket{1_B}\big],\\
\ket{s_{0}^3}_B&=\frac{1}{\sqrt{6}}\big[-c_1 \ket{1_B}+c_2\ket{2_B}\big],\\
\ket{s_{0}^4}_B &=\frac{1}{\sqrt{6}}\big[c_2  \ket{0_B}+c_0\ket{2_B}\big],\\
\ket{s_{0}^5}_B&=\frac{1}{\sqrt{6}}\big[c_2\ket{0_B}-c_0\ket{2_B}\big],\\
\ket{s_{0}^6}_B&=\frac{1}{\sqrt{6}}\big[c_2 \ket{1_B}+c_1\ket{2_B}\big],\\
\ket{s_{0}^7}_B&=\frac{1}{\sqrt{6}}\big[c_2 \ket{1_B}- c_1\ket{2_B}\big],\\
\ket{s_{0}^8}_B&:=\frac{1}{3\sqrt{2}}\big[-2c_0 \ket{0_B}+c_1\ket{1_B}+c_2\ket{2_B}\big], 
\end{align}
\label{eq10}
\end{subequations}
Using Eq.\eqref{eq7} it is straightforward to see that the measurement gates to be, 
\begin{subequations}
\begin{align}
\Lambda_0^0& :=\frac{1}{3}\big[\ket{0}_B\bra{0}+\ket{1}_B\bra{1}+\ket{2}_B\bra{2}\big], \\
\Lambda_0^1&:=\frac{1}{\sqrt{6}}\big[\ket{0}_B\bra{1}+\ket{1}_B\bra{0}\big], \\
\Lambda_0^2&:=\frac{1}{\sqrt{6}}\big[\ket{0}_B\bra{1}-\ket{1}_B\bra{0}\big], \\
\Lambda_0^3&:=-\frac{1}{\sqrt{6}}\big[\ket{1}_B\bra{1}+\ket{2}_B\bra{2}\big], \\
\Lambda_0^4&:=\frac{1}{\sqrt{6}}\big[\ket{0}_B\bra{2}+\ket{2}_B\bra{0}\big], \\
\Lambda_0^5&:=\frac{1}{\sqrt{6}}\big[\ket{0}_B\bra{2}-\ket{2}_B\bra{0}\big], \\
\lambda_0^6&:=\frac{1}{\sqrt{6}}\big[\ket{1}_B\ket{2}+\ket{2}_B\bra{1}\big], \\
\Lambda_0^7&:=\frac{1}{\sqrt{6}}\big[\ket{1}_B\bra{2}-\ket{2}_B\bra{1}\big], \\
\Lambda_{0}^8 &:=\frac{1}{3\sqrt{2}}\big[\ket{1}_B\bra{1}+\ket{2}_B\bra{2}-2\big(\ket{0}_B\bra{0}\big]. 
\end{align}
\label{eq11}
\end{subequations}
\\\\
\noindent 
Finally we are in position to decipher the teleported state from the composite state $\ket{\xi_{0}}_{A_1A_2B}$ as per augmented protocol. To complete teleportation, Bob requires the information of state $\ket{\Psi_{k}^{}}_{A_1A_2}$ from Alice, which she will send via a classical channel respecting causality. After receiving that information, he could ascertain the appropriate measurement gate to recover the state $\ket{\phi}_B$ at `B' using Eq.\eqref{eq7}. The non-unitarity of measurement operators, which arises plausibly due to the non-unitarity of the Gell-Mann matrices, is an important outcome of our treatment which differs significantly from the standard Bennett's protocol with EPR states, where the Pauli operators used as measurement operators are unitary. Table I summarizes the result of teleporting the qutrit state from Alice to Bob, i.e., from $\ket{\phi}_A$ to $\ket{\phi}_B$), for the shared state $\ket{\Psi_0}_{A_1B}$: 
\begin{table}[ht]
\centering
\caption{Teleportation table using shared entangled state $\ket{\Psi_{0}}_{A_2B}$ }
\label{tab:teleportation}
\begin{threeparttable}
\begin{tabular}{ccccc}
\toprule
Alice's channel & Pre-measurement & Measurement & Difference \tnote{1} & Remarks \\ ($\ket{\Psi_k}_{A_1A_2}$) & State  ($\ket{s^k_0}_B$)  & Gate ($\Lambda_0^k$) &  ($\Delta_{QT}(\ket{\Psi_0}_{A_1A_2})\big)$  \\
\midrule 
$\ket{\Psi^{}_{0}}_{A_1A_2}$ & $\ket{s^0_{0}}_B$ & $\Lambda_0^0$ & $0$ & Unchanged\\ 
$\ket{\Psi^{}_{1}}_{A_1A_2}$ & $\ket{s^1_{0}}_B$ & $\Lambda_{0}^1$ & $0$ & Teleported\\
$\ket{\Psi^{}_{2}}_{A_1A_2}$  & $\ket{s^{2}_{0}}_B$ & $\Lambda_{0}^2$ & $0$ & Teleported\\
$\ket{\Psi^{}_{3}}_{A_1A_2}$  & $\ket{s^3_{0}}_0$ & $\Lambda_0^3$ & 0 & Teleported\\ 
$\ket{\Psi^{}_{4}}_{A_1A_2}$ & $\ket{s^4_{0}}_B$ & $\Lambda_{0}^{4}$ & 0 & Teleported\\
$\ket{\Psi^{}_{5}}_{A_1A_2}$ & $\ket{s^5_{0}}_B$ & $\Lambda_{0}^5$ & 0 & Teleported\\
$\ket{\Psi_{6}^{}}_{A_1A_2}$ & $\ket{s^6_{0}}_B$ & $\Lambda_{0}^{6}$ & 0 & Teleported\\
$\ket{\Psi^{}_{7}}_{A_1A_2}$  & $\ket{s^7_{0}}_B$ & $\Lambda_{0}^7$ & 0 & Teleported\\
$\ket{\Psi^{}_{8}}_{A_1A_2}$ & $\ket{s^8_{0}}_B$ & $\Lambda_0^8$ & 0 & Teleported\\
\bottomrule
\end{tabular}
\begin{tablenotes}
\item[1] $\Delta_{QT}\big({\ket{\Psi_{k}}_{A_1A_2}}\big):=\ket{s^{k}_i}_B-\big[\Lambda_i^k\big]\ket{\phi}_B$
\end{tablenotes}
\end{threeparttable}
\end{table}

\noindent
The teleportation using remaining channels is similar, and the results are presented in the Appendix.

\section{Conclusion}
\par
In this paper, we investigate quantum teleportation of a qutrit system using all possible two-qutrit entangled states permitted by SU(3) symmetry and derive the measurement gates for each teleportation channel. 
These measurement gates are found to be non-unitary, unlike the unitary gates used in the standard protocol, and therefore necessitate a minimal modification of Bennett's protocol. The non-unitarity suggests the potential for implementing Measurement-Based Quantum Computing (MBQC) with cluster-state based robust network for qutrit systems in higher dimension, rather than relying on conventional unitary gate-based circuit \cite{Raussendorf2001,Raussendorf2003,Nielsen2006}. Apart from that, the non-unitarity indicates the presence of certain dissipative effects arising from environmental interactions, measurement-induced state collapse, or other sources of noise, the extent of which varies across teleportation channels 
\cite{Oh2002, Sebastian2023, Liu2024}. A more detailed study of SU(3)-based entangled states is necessary to enhance the fidelity, scalability, and security of next-generation communication system with complex quantum networks.

\pagebreak

\pagebreak 

\begin{center}
    {\LARGE \textbf{APPENDIX}}
\end{center}

In this Appendix, we have charted the pre-measurement states and the measurement gates involved in teleporting the remaining qutrit state using different shared entangled states: 

\section*{\LARGE \textbf{i) Teleportation using $\ket{\Psi_{1}}_{A_2B}$ state: }}

For channel II, the pre-measurement states at B are given by, 
\begin{subequations}
\begin{align}
\ket{s_{1}^0}_B&:=\frac{1}{\sqrt{6}}\big[c_0\ket{1_B}+c_1\ket{0_B}\big],\\
\ket{s_{1}^1}_B&=\frac{1}{2}\big[c_0 \ket{0_B} + c_1\ket{1_B}\big],\\
\ket{s_{1}^2}_B&=\frac{1}{2}\big[-c_0 \ket{0_B} + c_1\ket{1_B}\big],\\
\ket{s_{1}^3}_B&=-\frac{1}{2}c_1\ket{0_B},\\
\ket{s_{1}^4}_B &=\frac{1}{2}c_2\ket{1_B},\\
\ket{s_{1}^5}_B&=\frac{1}{2}c_2\ket{1_B},\\
\ket{s_{1}^6}_B&=\frac{1}{2}c_2\ket{0_B},\\
\ket{s_{1}^7}_B&=\frac{1}{2}c_2\ket{0_B},\\
\ket{s_{1}^8}_B&:=\frac{1}{2\sqrt{3}}\big[c_1 \ket{0_B}-2c_0\ket{1_B}\big], 
\end{align}
\end{subequations}
and corresponding measurement gates are,
\begin{subequations}
\begin{align}
\hat{\Lambda}_1^0& := \frac{1}{\sqrt{6}}\ket{0}\bra{1}+\ket{1}\bra{0}, \\
\hat{\Lambda}_1^1&:= \frac{1}{2}\ket{0}\bra{0}+\ket{1}\bra{1}, \\
\hat{\Lambda}_1^2&:=-\frac{1}{2}\ket{0}\bra{0}+\ket{1}\bra{1}, \\
\hat{\Lambda}_1^3&:= -\frac{1}{2}\ket{0}\bra{1}, \\
\hat{\Lambda}_1^4&:=\frac{1}{2}\ket{1}\bra{2}, \\
\hat{\Lambda}_1^5&:=\frac{1}{2}\ket{1}\bra{2}, \\
\hat{\Lambda}_1^6&:=\frac{1}{2}\ket{0}\bra{2}, \\
\hat{\Lambda}_1^7&:=\frac{1}{2}\ket{0}\bra{2}, \\
\hat{\Lambda}_{1}^8&:=\frac{1}{2\sqrt{3}}\big[\ket{0}\bra{1}-2\ket{1}\bra{0}.
\end{align}
\end{subequations}

\pagebreak 

\section*{\LARGE \textbf{ii) Teleportation using $\ket{\Psi_{2}}_{A_2B}$ state: }}

For channel III, the pre-measurement states at site `B' are given by,
\begin{subequations}
\begin{align}
\ket{s_{2}^0}_B&:=\frac{1}{\sqrt{6}}\big[c_1 \ket{0_B}-c_0\ket{1_B}\big],\\
\ket{s_{2}^1}_B&:=\frac{1}{2}\big[c_0\ket{0_B}-c_1\ket{1_B}\big],\\
\ket{s_{2}^2}_B&:=-\frac{1}{2}\big[c_0\ket{0_B}+c_1\ket{1_B}\big],\\
\ket{s_{2}^3}_B&:=-\frac{1}{2}c_1\ket{0_B},\\
\ket{s_{2}^4}_B&:=-\frac{1}{2}c_2\ket{1_B},\\
\ket{s_{2}^5}_B&:=-\frac{1}{2}c_2\ket{1_B},\\
\ket{s_{2}^6}_B&:= \frac{1}{2}c_2\ket{0_B},\\
\ket{s_{2}^7}_B&:= \frac{1}{2}c_2\ket{0_B},\\
\ket{s_{2}^8}_B&:=\frac{1}{2\sqrt{3}}\big[2c_0\ket{1_B}+c_1\ket{0_B}\big]. 
\end{align}
\end{subequations}
and corresponding measurement gates are:
\begin{subequations}
\begin{align}
\Lambda_2^0& := \frac{1}{\sqrt{6}}\big[\ket{0}\bra{1}-\ket{1}\bra{0}\big], \\
\Lambda_2^1&:=\frac{1}{2}\ket{0}\bra{0}-\ket{1}\bra{1}, \\
\Lambda_2^2&:=-\frac{1}{2}\ket{0}\bra{0}-\ket{1}\bra{1},\\
\Lambda_2^3&:= -\frac{1}{2}\ket{0}\bra{1},\\
\Lambda_2^4&:=-\frac{1}{2}\ket{1}\bra{2},\\
\Lambda_2^5&:=-\frac{1}{2}\ket{1}\bra{2},\\
\Lambda_2^6&:=\frac{1}{2}\ket{0}\bra{2},\\
\Lambda_2^7&:=\frac{1}{2}\ket{0}\bra{2},\\
\Lambda_{2}^8&:=\frac{1}{2\sqrt{3}}\big[2\ket{1}\bra{0}+\ket{0}\bra{1}\big].
\end{align}
\end{subequations}

\pagebreak 

\section*{\LARGE \textbf{iii) Teleportation using $\ket{\Psi_{3}}_{A_2B}$ state: }}

For channel IV, the pre-measurement states at site `B' are given by,
\begin{subequations}
\begin{align} 
\ket{s_{3}^0}_B&:=\frac{1}{\sqrt{6}}\big[-c_1\ket{1_B}+c_2\ket{2_B}\big]\\
\ket{s_{3}^1}_B&:= -\frac{1}{2}c_0\ket{1_B}\\
\ket{s_{3}^2}_B&:= \frac{1}{2}c_0\ket{1_B},\\
\ket{s_{3}^3}_B&:=\frac{1}{2}\big[c_1\ket{1_B}+c_2\ket{2_B}\big],\\
\ket{s_{3}^4}_B&:=\frac{1}{2}c_0\ket{2_B} \\ 
\ket{s_{3}^5}_B&:=-\frac{1}{2}c_0\ket{2_B}\\
\ket{s_{3}^6}_B&:=\frac{1}{2}\big[c_1 \ket{2_B}-c_2\ket{1}_B\big]\\
\ket{s_{3}^7}_B&:=-\frac{1}{2}\big[c_1 \ket{2}_B+c_2\ket{1_B}_B\big],\\
\ket{s_{3}^8}_B&:=\frac{1}{2\sqrt{3}}\big[-c_1 \ket{1_B}+c_2\ket{2_B}\big]. 
\end{align}
\end{subequations}
and the measurement gates and the gates are given by:
\begin{subequations}
\begin{align}
\Lambda_3^0& :=\frac{1}{\sqrt{6}}\big[-\ket{1}\bra{1}+\ket{2}\bra{2}\big] \\
\Lambda_3^1&:=-\frac{1}{2}\ket{1}\bra{0}\\
\Lambda_3^2&:=\frac{1}{2}\ket{1}\bra{0}\\
\Lambda_3^3&:=\frac{1}{2}\big[\ket{1}\bra{1}+\ket{2}\bra{2}\big] \\
\Lambda_3^4&:=\frac{1}{2}\ket{2}\bra{0} \\
\Lambda_3^5&:= -\frac{1}{2}\ket{2}\bra{0}\\
\Lambda_3^6&:=\frac{1}{2}\big[-\ket{1}\bra{2}+\ket{2}\bra{1}\big] \\
\Lambda_3^7&:=-\frac{1}{2}\big[\ket{1}\bra{2}+\ket{2}\bra{1}\big]\\
\Lambda_{3}^8&:=\frac{1}{2\sqrt{3}}\big[-\ket{1}\bra{1}+\ket{2}\bra{2}\big].
\end{align}
\end{subequations}

\pagebreak

\section*{\LARGE \textbf{iv) Teleportation using $\ket{\Psi_{4}}_{A_2B}$ state:}}
\par 
For channel V, the pre-measurement states at site `B' are given by, 
\begin{subequations}
\begin{align}
\ket{s_{4}^0}_B&:=\frac{1}{\sqrt{6}}\big[ c_0 \ket{2_B}+c_2\ket{0_B}\big],\\
\ket{s_{4}^1}_B&:=\frac{1}{2}c_1\ket{2_B},\\
\ket{s_{4}^2}_B&:=\frac{1}{2}c_1\ket{2_B},\\
\ket{s_{4}^3}_B&:=\frac{1}{2}c_2\ket{0_B},\\
\ket{s_{4}^4}_B&:=\frac{1}{2}\big[c_2\ket{2}_B+c_0\ket{0_B}\big],\\
\ket{s_{4}^5}_B&:=\frac{1}{2}\big[c_2\ket{2_B}-c_0\ket{0_B}\big],\\
\ket{s_{4}^6}_B&:=\frac{1}{2}c_1\ket{0_B},\\
\ket{s_{4}^7}_B&:=-\frac{1}{2}c_1\ket{0_B},\\
\ket{s_{4}^8}_B&:=\frac{1}{2\sqrt{3}}\big[c_2\ket{0_B}-2c_0\ket{2_B}\big]. 
\end{align}
\end{subequations}
and corresponding measurement gates are, 
\begin{subequations}
\begin{align}
\Lambda_4^0& := \frac{1}{\sqrt{6}}\big[\ket{0}\bra{2}+\ket{2}\bra{0}\big], \\
\Lambda_4^1&:=\frac{1}{2}\ket{2}\bra{1}, \\
\Lambda_4^2&:=\frac{1}{2}\ket{2}\bra{1}, \\
\Lambda_4^3&:=\frac{1}{2}\ket{0}\bra{2}, \\
\Lambda_4^4&:=\frac{1}{2}\big[\ket{2}\bra{2}+\ket{0}\bra{0}\big], \\
\Lambda_4^5&:=\frac{1}{2}\big[\ket{2}\bra{2}-\ket{0}\bra{0}\big], \\
\Lambda_4^6&:=\frac{1}{2}\ket{0}\bra{1}, \\
\Lambda_4^7&:=-\frac{1}{2}\ket{0}\bra{1},\\
\Lambda_{4}^8&:=\frac{1}{2\sqrt{3}}\big[\ket{0}\bra{2}-2\ket{2}\bra{0}\big].
\end{align}
\end{subequations}

\pagebreak 

\section*{\LARGE \textbf{v) Teleportation using $\ket{\Psi_{5}}_{A_2B}$ state: }}
\par 
For channel VI, the pre-measurement states at site `B' are given by, 
\begin{subequations}
\begin{align}
\ket{s_{5}^0}_B&=\frac{1}{\sqrt{6}}\big[c_2 \ket{0_B}-c_0\ket{2_B}\big],\\
\ket{s_{5}^1}_B&=-\frac{1}{2}c_1 \ket{2_B},\\
\ket{s_{5}^2}_B&=-\frac{1}{2}c_1 \ket{2_B},\\
\ket{s_{5}^3}_B&=\frac{1}{2}c_2 \ket{0_B} ,\\
\ket{s_{5}^4}_B&=\frac{1}{2}\big[c_0\ket{0_B}-c_2\ket{2_B}\big],\\
\ket{s_{5}^5}_B&=-\frac{1}{2}\big[c_0\ket{0_B}+c_2\ket{2_B}\big],\\
\ket{s_{5}^6}_B&=\frac{1}{2}c_1 \ket{0_B},\\
\ket{s_{5}^7}_B&=-\frac{1}{2}c_1 \ket{0_B},\\
\ket{s_{5}^8}_B&:=\frac{1}{2\sqrt{3}}\big[2 c_0 \ket{2_B}+c_2\ket{0_B}\big],
\end{align}
\end{subequations}
and corresponding measurement gates are,
\begin{subequations}
\begin{align}
\Lambda_5^0& :=\frac{1}{\sqrt{6}}\big[\ket{0}\bra{2}-\ket{2}\bra{0}\bigg],
\\
\Lambda_5^1&:=-\frac{1}{2}\ket{2}\bra{1},\\
\Lambda_5^2&:=-\frac{1}{2}\ket{2}\bra{1},\\
\Lambda_5^3&:=\frac{1}{2}\big[\ket{0}\bra{2}\\
\Lambda_5^4&:=\frac{1}{2}\big[\ket{0}\bra{0} - \ket{2}\bra{2} \big],\\
\Lambda_5^5&:=-\frac{1}{2}\big[\ket{0}\bra{0} + \ket{2}\bra{2} \big],\\
\Lambda_5^6&:=\frac{1}{2}\ket{0}\bra{1},\\
\Lambda_5^7&:=-\frac{1}{2}\ket{0}\bra{1},\\
\Lambda_{5}^8&:=\frac{1}{2\sqrt{3}}\big[\ket{0}\bra{2}+2\ket{2}\bra{0}\big]. 
\end{align}
\end{subequations}

\pagebreak 

\section*{\LARGE \textbf{vi) Teleporting using $\ket{\Psi_6}_{A_2B}$ state: }}
\par 
For channel VII, the pre-measurement states at site `B' are given by,
\begin{subequations}
\begin{align}
\ket{s_{6}^0}_B&=\frac{1}{\sqrt{6}}\big[c_1 \ket{0}_B+c_2\ket{1}_B\big], \\
\ket{s_{6}^1}_B&=\frac{1}{2}c_0 \ket{0_B},\\
\ket{s_{6}^2}_B&=-\frac{1}{2}c_0 \ket{0_B},\\
\ket{s_{6}^3}_B&=\frac{1}{2}\big[c_2 \ket{1_B} - c_1\ket{0_B}\big],\\
\ket{s_{6}^4}_B&=\frac{1}{2}c_0\ket{1_B},\\
\ket{s_{6}^5}_B&=-\frac{1}{2}c_0\ket{1_B}\\
\ket{s_{6}^6}_B&=\frac{1}{2}\big[c_2\ket{0_B}+c_1 \ket{1_B}\big],\\
\ket{s_{6}^7}_B&=-\frac{1}{2}\big[c_2\ket{0_B}-c_1 \ket{1_B}\big],\\
\ket{s_{6}^8}_B&:=\frac{1}{2\sqrt{3}}\big[c_2 \ket{1_B}+c_1\ket{0_B}\big].
\end{align}
\end{subequations}
and the corresponding measurement gates are,
\begin{subequations}
\begin{align}
\Lambda_6^0& := \frac{1}{\sqrt{6}}\big[\ket{0}\bra{1}+\ket{1}\bra{2}\big],\\
\Lambda_6^1&:=\frac{1}{2}\ket{0}\bra{0},\\
\Lambda_6^2&:=-\frac{1}{2}\ket{0}\bra{0},\\
\Lambda_6^3&:=\frac{1}{2}\big[-\ket{0}\bra{1}+\ket{1}\bra{2}\big],\\
\Lambda_6^4&:=\frac{1}{2}\big[-\ket{0}\bra{1} + \ket{1}\bra{2}\big],\\
\Lambda_6^5&:=-\frac{1}{2}\ket{1}\bra{0},\\ 
\Lambda_6^6&:=\frac{1}{2}\big[\ket{0}\bra{2}+\ket{1}\bra{1}\big],\\
\Lambda_6^7 &:=\frac{1}{2}\big[\ket{0}\bra{2}-\ket{1}\bra{1}\big],\\
\Lambda_{6}^8&:=\frac{1}{2\sqrt{3}}\big[\ket{0}\bra{1}+\ket{1}\bra{2}\big].
\end{align}
\end{subequations}

\pagebreak 

\section*{\LARGE \textbf{vii) Teleporting using $\ket{\Psi_7}_{A_2B}$ state: }}
\par 
For channel VIII, the pre-measurement states at site `B' are given by,
\begin{subequations}
\begin{align}
\ket{s_{7}^0}_B&=\frac{1}{\sqrt{6}}\big[c_2 \ket{1_B}-c_1\ket{0_B}\big],\\
\ket{s_{7}^1}_B&=-\frac{1}{2}c_0 \ket{0_B},\\
\ket{s_{7}^2}_B&=\frac{1}{2}c_0 \ket{0_B},\\
\ket{s_{7}^3}_B&=\frac{1}{2}c_1 \ket{0_B} + c_2\ket{1_B}\big],\\
\ket{s_{7}^4}_B&=\frac{1}{2}c_0\ket{1_B},\\
\ket{s_{7}^5}_B&=-\frac{1}{2}c_0\ket{1_B},\\
\ket{s_{7}^6}_B&=\frac{1}{2}\big[c_1 \ket{1_B}-c_2\ket{0_B}\big],\\
\ket{s_{7}^7}_B&=-\frac{1}{2}\big[c_1 \ket{1_B}+c_2\ket{0_B}\big],\\
\ket{s_{7}^8}_B&:=\frac{1}{2\sqrt{3}}\big[c_2 \ket{1_B}-c_1\ket{0_B}\big]
\end{align}
\end{subequations}\
and corresponding measurement gates are, 
\begin{subequations}
\begin{align}
\Lambda_7^0& := \frac{1}{\sqrt{6}}\big[-\ket{0}\bra{1}+\ket{1}\bra{2}\big],\\
\Lambda_7^1&:=-\frac{1}{2}\ket{0}\bra{0}, \\
\Lambda_7^2&:=\frac{1}{2}\ket{0}\bra{0}, \\
\Lambda_7^3&:=\frac{1}{2}\big[\ket{0}\bra{1}+\ket{1}\bra{2}\big], \\
\Lambda_7^4&:=\frac{1}{2}\ket{1}\bra{0}, \\
\Lambda_7^5&:=-\frac{1}{2}\ket{1}\bra{0}, \\
\Lambda_7^6&:=\frac{1}{2}\big[\ket{1}\bra{1} - \ket{0}\bra{2}\big],  \\
\Lambda_7^7&:=-\frac{1}{2}\big[\ket{1}\bra{1}+\ket{0}\bra{2}\big],\\
\Lambda_{7}^8&:=\frac{1}{2\sqrt{3}}\big[-\ket{0}\bra{1}+\ket{1}\bra{2}\big].
\end{align}
\end{subequations}

\pagebreak 

\section*{\LARGE \textbf{viii) Teleportation using $\ket{\Psi_{8}}_{A_2B}$ state: }}

For channel IX, the pre-measurement states at site `B' are given by,
\begin{subequations}
\begin{align}
\ket{s_{8}^0}_B&:=\frac{1}{3\sqrt{2}}\big[c_1\ket{1_B} - 2c_0\ket{2_B}\big],\\
\ket{s_{8}^1}_B&:=\frac{1}{2\sqrt{3}}\big[c_0 \ket{1_B}-2c_1\ket{0_B},\\
\ket{s_{8}^2}_B&:=-\frac{1}{2\sqrt{3}}\big[c_0 \ket{1_B}+2c_1\ket{0_B}\big],\\
\ket{s_{8}^3}_B&:=-\frac{1}{2\sqrt{3}}\big[c_1 \ket{1_B}\big],\\
\ket{s_{8}^8}_B&:=\frac{1}{2\sqrt{3}}\big[c_0\ket{2_B} - 2c_2\ket{0_B}\big],\\
\ket{s_{8}^4}_B&:=-\frac{1}{2\sqrt{3}}\big[c_0\ket{2_B} +2c_2\ket{0_B}\big],\\
\ket{s_{8}^5}_B&:=\frac{1}{2\sqrt{3}}\big[c_2\ket{1_B} +c_1\ket{2_B}\big],\\
\ket{s_{8}^7}_B&:=\frac{1}{2\sqrt{3}}\big[c_2\ket{1_B} -c_1\ket{2_B}\big],\\
\ket{s_{8}^8}_B&:=\frac{1}{6}\big[4 c_0 \ket{0_B}+c_1\ket{1_B}\big].
\end{align}
\end{subequations}
and corresponding measurement gates are, 
\begin{subequations}
\begin{align}
\Lambda_0^8& := \frac{1}{3\sqrt{2}}\big[-2\ket{0}\bra{0}+\ket{1}\bra{1}\big], \\
\Lambda_1^8&:=\frac{1}{2\sqrt{3}}\big[-2\ket{0}\bra{1}+\ket{1}\bra{0}\big],\\
\Lambda_2^8&:=-\frac{1}{2\sqrt{3}}\big[2\ket{0}\bra{1}+\ket{1}\bra{0}\big], \\
\Lambda_3^8&:=-\frac{1}{2\sqrt{3}}\ket{1}\bra{1}, \\
\Lambda_4^8&:=-\frac{1}{2\sqrt{3}}\big[-2\ket{0}\bra{2}+\ket{2}\bra{0}\big],\\
\Lambda_5^8&:=-\frac{1}{2\sqrt{3}}\big[2\ket{0}\bra{2}+\ket{2}\bra{0}\big],\\
\Lambda_6^8&:=\frac{1}{2\sqrt{3}}\big[\ket{1}\bra{2}+\ket{2}\bra{1}\big], \\
\Lambda_7^8&:=\frac{1}{2\sqrt{3}}\big[\ket{1}\bra{2}-\ket{2}\bra{1}\big], \\
\Lambda_{8}^8&:=\frac{1}{6}\big[4\ket{0}\bra{0}+\ket{1}\bra{1}\big].
\end{align}
\end{subequations}

\pagebreak 



\bibliographystyle{unsrt}
\bibliography{teleportation}

\begin{thebibliography}{10}

\bibitem{Bennett1993}
C.~H. Bennett, G.~Brassard, C.~Cr{\'e}peau, R.~Jozsa, A.~Peres, and W.~K. Wootters.
\newblock Teleporting an unknown quantum state via dual classical and einstein-podolsky-rosen channels.
\newblock {\em Physical Review Letters}, 70(13):1895--1899, 1993.

\bibitem{Nielsen2010}
M.~A. Nielsen and I.~L. Chuang.
\newblock {\em Quantum Computation and Quantum Information: 10th Anniversary Edition}.
\newblock Cambridge University Press, 2010.

\bibitem{Bouwmeester1997}
D.~Bouwmeester, J.~W. Pan, M.~Mattle, Ma. Eibl, H.~Weinfurter, and A.~Zeilinger.
\newblock Experimental quantum teleportation.
\newblock {\em Nature}, 390(6660):575--579, 1997.

\bibitem{Zeilinger2000}
A.~Zeilinger.
\newblock Quantum teleportation.
\newblock {\em Scientific American}, 282(4):50--59, 2000.

\bibitem{Bao2012}
X~Bao et.al.
\newblock Quantum teleportation between remote atomic-ensemble quantum memories.
\newblock {\em Proceedings of the National Academy of Sciences}, 109(50):20347--20351, 2012.

\bibitem{Boschi1998}
D.~Boschi, S.~Branca, De~Martini F., L.~Hardy, and S.~Popescu.
\newblock Experimental realization of teleporting an unknown pure quantum state via dual classical and einstein-podolsky-rosen channels.
\newblock {\em Physical Review Letters}, 80(6):1121--1125, 1998.

\bibitem{Valivarthi2016}
R.~Valivarthi, M.~G. Puigibert, Zhou S., A.~T. Knaut, V.~B. Verma, M.~D. Shaw, F.~Marsili, S.~D. Dyer, S.~Nam, D.~Oblak, and W.~Tittel.
\newblock Quantum teleportation across a metropolitan fibre network.
\newblock {\em Nature Photonics}, 10:776--780, 2016.

\bibitem{Shen2023}
Si~Shen et.al.
\newblock Hertz-rate metropolitan quantum teleportation.
\newblock {\em Light: Science \& Applications}, 12:115, 2023.

\bibitem{Bennett1984}
C.H. Bennett and G.~Brassard.
\newblock Quantum cryptography: Public key distribution and coin tossing.
\newblock In {\em Proceedings of the IEEE International Conference on Computers, Systems and Signal Processing}, pages 175--179, Bangalore, India, 1984. IEEE.

\bibitem{Gisin2002}
G.~Gisin, G.~Ribordy, W.~Tittel, and H.~Zbinden.
\newblock Quantum cryptography.
\newblock {\em Reviews of Modern Physics}, 74(1):145--195, 2002.

\bibitem{Takeda2013}
S.~Takeda, T.~Mizuta, M.~Fuwa, P~van Loock, and A.~Furusawa.
\newblock Deterministic quantum teleportation of photonic quantum bits by a hybrid technique.
\newblock {\em Nature}, 500(7462):315--318, August 2013.

\bibitem{DeMartini2019}
F.~De~Martini and F.~Sciarrino.
\newblock Twenty years of quantum state teleportation at the sapienza university in rome.
\newblock {\em Entropy}, 21(8):768, 2019.

\bibitem{Pirandola2015}
S.~Pirandola, J.~Eisert, C.~Weedbrook, A.~Furusawa, and S.~L. Braunstein.
\newblock Advances in quantum teleportation.
\newblock {\em Nature Photonics}, 9(10):641--652, 2015.

\bibitem{Hossain2024}
I.~H. Mohammad, A.~S. Shaharier, Md.~Hasan Habib, F~Akter, Badhon~M. B, and M.~N.~U. Islam.
\newblock Quantum-edge cloud computing: A future paradigm for iot applications.
\newblock {\em arXiv preprint \,}, 2024.
\newblock h.

\bibitem{Sikiru2024}
I.L.A. Sikiru, A.~D. Kora, E.~C. Ezin, L.~Imoize, Agbotiname, and Chun-Ta Li.
\newblock Hybridization of learning techniques and quantum mechanism for iiot security: Applications, challenges, and prospects.
\newblock {\em Electronics}, 13(21), 2024.

\bibitem{Hioe1982}
F.~T. Hioe and J.~H. Eberly.
\newblock Nonlinear constants of motion for three-level quantum systems.
\newblock {\em Physical Review A}, 25:2168--2183, 1982.

\bibitem{Yoo1985}
H.~J. Yoo and J.~H. Eberly.
\newblock Dynamical theory of an atom with two or three levels interacting with quantized cavity fields.
\newblock {\em Physics Reports}, 118(5-6):239--337, 1985.

\bibitem{Nath2008}
M.~R. Nath, S.~Sen, A.~K. Sen, and G.~Gangopadhyay.
\newblock Dynamical symmetry breaking of lambda and vee-type three-level systems on quantization of the field modes.
\newblock {\em Pramana - Journal of Physics}, 71:77--97, 2008.

\bibitem{Sen2012}
S.~Sen, M.~R. Nath, T.~K. Dey, and G.~Gangopadhyay.
\newblock Bloch space structure, the qutrit wave function and atom–field entanglement in three-level systems.
\newblock {\em Annals of Physics}, 327:224--252, 2012.

\bibitem{Dicke1954}
R.~H. Dicke.
\newblock Coherence in spontaneous radiation processes.
\newblock {\em Physical Review}, 93(1):99--110, 1954.

\bibitem{Misra1977}
B.~Misra and E.~C.~G. Sudarshan.
\newblock The zeno's paradox in quantum theory.
\newblock {\em Journal of Mathematical Physics}, 18(4):756--763, 1977.

\bibitem{Alzetta1976}
G.~Alzetta, A.~Gozzini, L.~Moi, and G.~Orriols.
\newblock Experimental method to obtain atomic coherence at radiofrequencies. application to level crossing resonance in optical pumping.
\newblock {\em Nuovo Cimento B}, 36(1):5--20, 1976.

\bibitem{Gray1978}
H.~R. Gray, R.~M. Whitley, and C.~R. Stroud.
\newblock Coherent trapping of atomic populations.
\newblock {\em Optics Letters}, 3(6):218--220, 1978.

\bibitem{Harris1990}
S.~E. Harris, J.~E. Field, and A.~Imamo{\u{g}}lu.
\newblock Nonlinear optical processes using electromagnetically induced transparency.
\newblock {\em Physical Review Letters}, 64(10):1107--1110, 1990.

\bibitem{Boller1991}
K.-J. Boller, A.~Imamo{\u{g}}lu, and S.~E. Harris.
\newblock Observation of electromagnetically induced transparency.
\newblock {\em Physical Review Letters}, 66(20):2593--2596, 1991.

\bibitem{Harris1989}
S.~E. Harris.
\newblock Lasers without inversion: Interference of lifetime-broadened resonances.
\newblock {\em Physical Review Letters}, 62(9):1033--1036, 1989.

\bibitem{Sen2015}
S.~Sen, T.~K. Dey, M.~R. Nath, and G.~Gangopadhyay.
\newblock Comparison of electromagnetically induced transparency in lambda, cascade and vee three-level systems.
\newblock {\em Journal of Modern Optics}, 62(2):166--174, 2015.

\bibitem{Sen2017}
S.~Sen, T.~K. Dey, and B.~Deb.
\newblock A unified approach to $\lambda$, v and $\xi$-type systems with one continuum.
\newblock {\em Journal of Modern Optics}, 64:2083--2096, 2017.

\bibitem{Sen2023}
S.~Sen, T.~K. Dey, and B.~Deb.
\newblock Resonance fluorescence in $\lambda$, v, and $\xi$-type three-level configurations.
\newblock {\em Physica Scripta}, 98(11):115124, 2023.

\bibitem{Lanyon2009}
et.~al. Benjamin P.~Lanyon.
\newblock Simplifying quantum logic using higher-dimensional {H}ilbert spaces.
\newblock {\em Nature Physics}, 5(2):134--140, Feb 2009.

\bibitem{PanLu2006}
F.~Pan and G.~Lu.
\newblock Classification and quantification of entangled bipartite qutrit pure states.
\newblock {\em International Journal of Modern Physics B}, 20(10):1333--1342, 2006.

\bibitem{Luo2019}
Yi-Han~Luo et. al.
\newblock Quantum teleportation in high dimensions.
\newblock {\em Physical Review Letters}, 123:070505, 2019.

\bibitem{Erhard2020}
M.~Erhard, M.~Krenn, and A.~Zeilinger.
\newblock Advances in high-dimensional quantum entanglement.
\newblock {\em Nature Reviews Physics}, 2(7):365--381, 2020.

\bibitem{Huang2020}
Y.~Huang and W.~Yang.
\newblock Quantum teleportation via qutrit entangled state.
\newblock {\em Chinese Journal of Electronics}, 29(2):228--232, 2020.

\bibitem{Leslie2019}
N.~Leslie, Devin J., and T.~W. Lynn.
\newblock Maximal lelm distinguishability of qubit and qutrit bell states using projective and non-projective measurements.
\newblock {\em arXiv preprint arXiv:1903.02655}, 2019.

\bibitem{Hu2020}
Xiao-Min et.~al. Hu.
\newblock Experimental high-dimensional quantum telepotation.
\newblock {\em Physical Review Letters}, 125:230501, 2020.

\bibitem{Roy2025}
S.~Roy, A.~Bhattacharjee, T.~K. Dey, and S.~Sen.
\newblock Unitary and non-unitary operators leverage perfect and imperfect single qutrit teleportation.
\newblock {\em arXiv preprint arXiv:2503.24247}, 2025.

\bibitem{Sen2024}
S.~Sen and T.~K. Dey.
\newblock An inequality for entangled qutrits in su (3) basis.
\newblock {\em Quantum Information Processing}, 23(7):267, 2024.

\bibitem{Raussendorf2001}
R.~Raussendorf and H.~J. Briegel.
\newblock A one-way quantum computer.
\newblock {\em Physical Review Letters}, 86:5188--5191, 2001.

\bibitem{Raussendorf2003}
R.~Raussendorf, D.~E. Browne, and H.~J. Briegel.
\newblock Measurement-based quantum computation on cluster states.
\newblock {\em Physical Review A}, 68(2):022312, August 2003.

\bibitem{Nielsen2006}
Michael~A. Nielsen.
\newblock Cluster-state quantum computation.
\newblock {\em Reports on Mathematical Physics}, 57(1):147--161, 2006.

\bibitem{Oh2002}
O.~Sangchul, L.~Soonchil, and L.~Hai-Woong.
\newblock Fidelity of quantum teleportation through noisy channels.
\newblock {\em Physical Review A}, 66(2):022316, 2002.

\bibitem{Sebastian2023}
A.~Sebastian, A.~N. Mansar, and N.~C. Randeep.
\newblock Beyond qubits: An extensive noise analysis for qutrit quantum teleportation.
\newblock {\em International Journal of Theoretical Physics}, 62(12):258, 2023.

\bibitem{Liu2024}
et.al. Zhao-Di~Liu.
\newblock Overcoming noise in quantum teleportation with multipartite hybrid entanglement.
\newblock {\em Science Advances}, 10(15):eadj3435, apr 2024.

\end{thebibliography}

\end{document}